# A NEW METHOD FOR THE DETERMINATION OF THE HUBBLE PARAMETER

SIDNEY VAN DEN BERGH

Dominion Astrophysical Observatory

5071 West Saanich Road

Victoria, British Columbia

V8X 4M6, Canada

## ABSTRACT

By chance, the slope of the $M_V$ (max) versus $(B-V)_{max}$ relation for recent theoretical models of supernovae of Type Ia (SNe Ia) by Höflich & Khokhlov is indistinguishable from the slope of a reddening line in the V versus B-V plane. This coincidence allows one to determine a parameter $M_V^*$ (max) for SNe Ia that is independent of both supernova detonation model and of interstellar reddening. Calibrating $M_V^*$ (max) with observations of SNe Ia by Hamuy et al. yields values of the Hubble parameter $H_o$ in the range 55 - 60 km s$^{-1}$ Mpc$^{-1}$. The discrepancy between this result, and values of $H_o$ recently obtained from observations of Cepheids in the Virgo cluster, suggests that either (1) the Cepheid distance scale is wrong, (2) the SN Ia models of Höflich & Khokhlov are too bright by ~ 0.75 mag near maximum light, or (3) their models are too red by ~ 0.25 mag in B-V.

Subject headings: Cosmology: distance scale - supernovae: general

1.   **INTRODUCTION**

Because of their high luminosities, supernovae of Type Ia would appear to be ideal calibrators of the extragalactic distance scale (Kowal 1968). However, observations collected during the last decade clearly show that SNe Ia are <u>not</u> all identical and that some of them have luminosities that deviate significantly from the mean. For example, SN 1957B and SN 1991bg (which both occurred in the almost dust-free elliptical M84) had luminosities at maximum light that differed by $\Delta$ B(max) $\simeq$ 2.4 mag. A second difficulty is that the majority of SNe Ia occur in dusty late-type spiral galaxies, in which the foreground reddening of individual supernovae is not known. In this <u>letter</u>, a method is suggested for circumventing both of these difficulties.

2.   **A REDDENING-FREE PARAMETER**

Recently, Höflich & Khokhlov (1995) have published a set of models for SNe Ia which encompass all currently discussed explosion scenarios. Their calculations include models for deflagrations, detonations, delayed detonations, pulsating delayed detonations and tamped detonations of Chandrasekhar mass progenitors, and helium detonations of low-mass white dwarfs. From their models, Höflich & Khokhlov construct multi-color light curves which are based on calculations that implicitly include radiation transport, expansion opacities, Monte-Carlo γ-ray transport, and molecule and dust formation. For some of these models, detailed non-LTE calculations were used. Figure 1 shows a plot of $M_V$ (max) versus $(B-V)_{max}$ for each of these models. The most interesting feature shown by this figure is that the faintest supernovae are also the reddest. This is in agreement with observations which show (Hamuy et al. 1994) that intrinsically faint SNe Ia, such as SN 1991bg and SN 1992K, were quite red at maximum light. Also shown in Fig. 1 is a reddening vector (Mathis 1990) with a slope $A_V = 3.1\ E(B-V)$. The fact that a reddening line with slope ≈ 3.1 fits the models for supernovae at maximum light implies that one can define a parameter

$$M_V^* (\text{max}) = M_V (\text{max}) - 3.1\ (B-V)_{max} \qquad (1)$$

which is, in first approximation, independent of both reddening <u>and</u> supernova model. For the 36 supernova models of Höflich & Khokhlov (1995) one find

$$M_V^* (\text{max}) = -19.60 \pm 0.05, \qquad (2)$$

with a dispersion of σ = 0.29 mag. For 13 supernovae for which Hamuy et al. (1995) have obtained detailed CCD-based light curves, one obtains

$$< M_V^*(\max) > - 5 \log (H_o / 85) = -18.83 \pm 0.11 \qquad (3)$$

Substituting Eqn. (2) into Eqn. (3) yields $H_o = 59.6 \pm 3.3$ km s$^{-1}$ Mpc$^{-1}$. Clearly, the 36 models of SNe Ia calculated by Höflich & Khokhlov do not populate the entire universe of possible supernova models. An alternative approach is to use only those models that give acceptable fits to light curves of reasonably well-observed supernovae of Type Ia. On average, each of the 26 best-observed light curves of SNe Ia, that are listed in Table 3 of Höflich & Khokhlov (1995), is fit reasonably well by about three different supernova models. For the 17 supernova models that fit at least one light curve well $< M_V^*(\max) > = -19.70 \pm 0.06$, with a dispersion of $\sigma = 0.24$ mag. The corresponding value of the Hubble parameter obtained from Eqn. (3) is $H_o = 56.9 \pm 3.2$ km s$^{-1}$ Mpc$^{-1}$. Finally, one can weight $M_V^*(\max)$ according to the number of well-observed light curves that are reasonably well fit by any one of the model light curves calculated by Höflich & Khokhlov. This approach yields $< M_V^*(\max) > = -19.75 \pm 0.02$, from which $H_o = 54.6 \pm 2.8$ km s$^{-1}$ Mpc$^{-1}$.

In summary, a comparison between the observed and computed light curves of SNe Ia at maximum light yields values of the Hubble parameter $H_o$ in the range 55-60 km s$^{-1}$ Mpc$^{-1}$. Note that these values of the Hubble parameter depend only on the input physics used by Höflich & Khokhlov, and are therefore independent of any astronomical calibration.

The values of $H_o$ obtained above are in reasonable agreement with $H_o$ (B light) = 50 ± 8 km s$^{-1}$ Mpc$^{-1}$ and $H_o$ (V light) = 54 ± 8 km s$^{-1}$ Mpc$^{-1}$ that Sandage & Tammann (1994) obtained from SNe Ia and the value $H_o$ = 67 ± 7 km s$^{-1}$ Mpc$^{-1}$ that Riess, Press & Kirshner (1995) find from light curve shapes. However, they disagree with many other determinations of the distance scale, such as those reviewed by Jacoby et al. (1992) and van den Bergh (1994).

In a recent paper, Hamry et al. (1995) presented detailed photometric observations of a small sample of 13 supernovae of Type Ia. After correcting for Galactic absorption (but not for internal absorption in the supernova's parent galaxy), their data yield an (unweighted) dispersion σ = 0.37 mag around a V(max) versus log cz relation of slope 5. Perhaps surprisingly, the present sample yields a slightly <u>larger</u> dispersion of σ = 0.40 mag[1] if the observed magnitudes are corrected by taking into account the maximum magnitude versus rate-of-decline relation adopted by Phillips (1993). The same dispersion $\sigma = 0.40$ mag is obtained from a Hubble diagram in which $V^*$ (max) / V - 3.1 (B-V) $_{max}$ is plotted versus log cz. The puzzling observation that reddening corrections to individual SNe Ia do not appear to reduce the dispersion in their Hubble diagram has previously been noted by van den Bergh (1993). Clearly, a larger sample of well-observed SNe Ia will be needed to draw conclusions that are not affected by the vagaries of small-number statistics.

---

[1] For the present small sample of SNe Ia, one obtains a lower dispersion σ = 0.27 mag if one adopts a slope that is only half as large as that derived by Phillips (1993)

**3.    DISCUSSION**

There are now four Cepheid-based distances to spiral galaxies in the Virgo region. For NGC 4321 (M 100), Freedman et al. (1994) obtained $(m-M)_o = 31.16 \pm 0.20$. More recently, Saha et al. (Saha 1995) have found $(m-M)_o = 31.10 \pm 0.15$ for NGC 4496 and $(m-M)_o = 31.05 \pm 0.15$ for NGC 4536. Finally, Pierce et al. (1994) derived $(m-M)_o = 30.91 \pm 0.15$ from ground-based CFHT observations of three Cepheids in NGC 4571. These remarkably consistent observations yield a weighted mean value $< (m-M)_o > = 31.04$ with a formal mean error of 0.08 mag. Allowing for a systematic error of 0.1 mag in the photometric calibrations of the Hubble Space Telescope (HST) photometry and an additional uncertainty of $^c$ 0.1 mag in the distance modulus of the Large Magellanic Cloud, one finally obtains a Virgo distance modulus (m-M)$_o$ = 31.04 ± 0.2 corresponding to a distance of 16.1 ± 1.5 (m.e.) Mpc. Combining this value with the well-determined D (Coma) / D (Virgo) value (van den Bergh 1992) and the Coma cluster velocity corrected to the rest frame of the microwave background (cf. Pierce et al. 1994) yields H$_o$ = 80 ± 8 km s$^{-1}$ Mpc$^{-1}$. This result is clearly inconsistent with the values of H$_o$ in the range 55-60 km s$^{-1}$ Mpc$^{-1}$ obtained in this <u>letter</u> by comparing SNe Ia observations and the light curves of Höflich & Khokhlov (1995). How can one account for this discrepancy? Possible explanations include the following possibilities: (1) the models of Höflich & Khokhlov yield SN Ia absolute magnitudes that are ~ 0.75 mag too bright, (2) the SN Ia models of Höflich & Khokhlov are ~ 0.25 mag too red, at maximum light, or (3) the zero-point of the Period-Luminosity relation for Cepheids is ~ 0.75 mag brighter than the one adopted by Madore & Freedman (1991). However, a comparison between the Period-Luminosity relation adopted by Madore & Freedman, and its recent calibration via Cepheids in clusters and associations (Laney & Stobie 1994), appears to rule out such a large error in the Cepheid calibration. For a recent comparison between Cepheid and other distance scales, the reader is referred to Mould et al. (1995).

I am indebted to Peter Höflich for providing me with corrections to a few of the tabulated data in his preprint, and to an anonymous referee for a number of helpful suggestions. Thanks are also due to Abi Saha for permission to quote new HST observations of Cepheids in NGC 4496 and NGC 4536.

**FIGURE LEGEND**

Fig. 1   $M_V$ (max) versus $(B-V)_{max}$ for 36 supernova models calculated by Höflich & Khokhlov.  Intrinsically faint supernovae are seen to have redder colors than luminous ones.  The arrow is a reddening line with slope $\Delta V = 3.1\ \Delta(B-V)$.

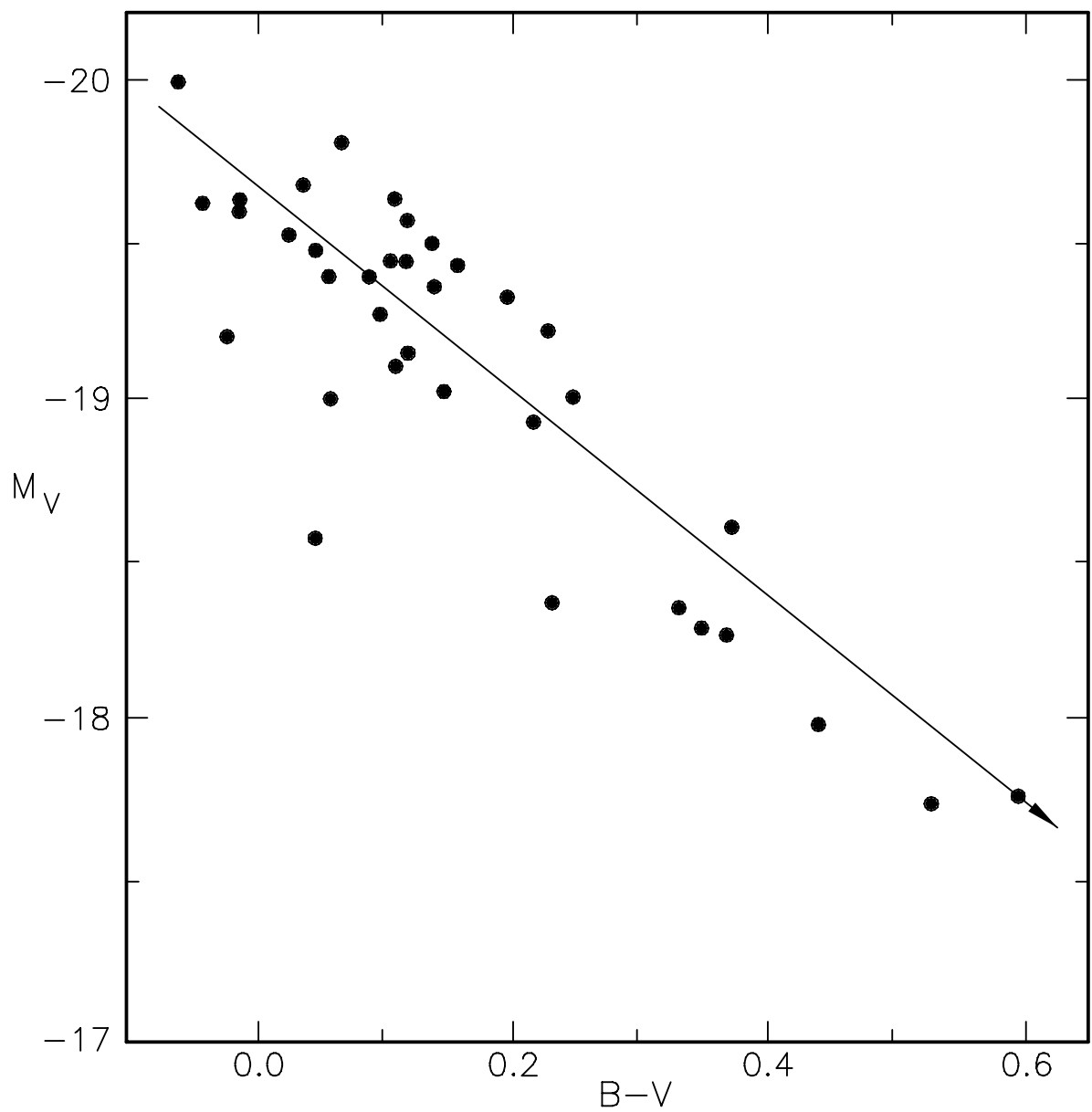